\documentstyle[12pt]{article}
\begin{document}
\newcommand{\Si}{\Sigma}
\newcommand{\tr}{{\rm tr}}
\newcommand{\ad}{{\rm ad}}
\newcommand{\Ad}{{\rm Ad}}
\newcommand{\ti}[1]{\tilde{#1}}
\newcommand{\om}{\omega}
\newcommand{\Om}{\Omega}
\newcommand{\de}{\delta}
\newcommand{\al}{\alpha}
\newcommand{\te}{\theta}
\newcommand{\ze}{\zeta}
\newcommand{\vth}{\vartheta}
\newcommand{\be}{\beta}
\newcommand{\lm}{\lambda}
\newcommand{\La}{\Lambda}
\newcommand{\D}{\Delta}
\newcommand{\ve}{\varepsilon}
\newcommand{\vf}{\varphi}
\newcommand{\G}{\Gamma}
\newcommand{\ip}{\hat{\upsilon}}
\newcommand{\Ip}{\hat{\Upsilon}}
\newcommand{\ga}{\gamma}
\newcommand{\li}{\lim_{n\rightarrow \infty}}
\newcommand{\si}{\sigma}
\newcommand{\beq}[1]{\begin{equation}\label{#1}}
\newcommand{\eq}{\end{equation}}
\newcommand{\beqn}[1]{\begin{eqnarray}\label{#1}}
\newcommand{\eqn}{\end{eqnarray}}
\newcommand{\p}{\partial}
\newcommand{\di}{{\rm diag}}
\newcommand{\rar}{\rightarrow}
\newcommand{\su}{{\bf su_2}}
\newcommand{\uo}{{\bf u_1}}
\newcommand{\ep}{\epsilon}
\newcommand{\GL}{{\rm GL}(N,{\bf C})}
\newcommand{\gl}{gl(N,{\bf C})}
\begin{flushright}
SISSA Ref. 125/95/EP\\
 ITEP-TH3/95\\
\end{flushright}
\vspace{10mm}
\begin{center}
{\Large\bf Generalized Hitchin systems and }\\
 {\Large\bf Knizhnik-Zamolodchikov-Bernard equation }\\
{\Large\bf on  elliptic curves }\\
\vspace{5mm}
M.A.Olshanetsky
\footnote{Research partially supported by ISF-MIF-300 , RFFI-94-02-14365
 and INTAS-94-2317 }
 \footnote{Talk given at 6 International Conference On Mathematical Physics,
  Moscow, June 1995} \\
{\sf SISSA, via Beirut 2-4, 34013 Trieste-Italy}\\
and\\
{\sf ITEP, 117259 Moscow, Russia,} \\
{\em e-mail olshanez@vxdesy.desy.de}\\
\vspace{5mm}
\end{center}
\begin{abstract}
Knizhnik-Zamolodchikov-Bernard (KZB) equation on an elliptic curve with
a marked point is derived by the classical Hamiltonian reduction and
further quantization. We consider  classical Hamiltonian systems on cotangent
 bundle to the loop
group $L(GL(N,{\bf C}))$ extended by the shift operators, to be related to the
elliptic
module. After the reduction we obtain the Hamiltonian system on cotangent
bundle to the  moduli of holomorphic principle bundles and the elliptic module.
It is a particular example of  generalized Hitchin systems (GHS)
 which are defined
as  hamiltonian systems on cotangent
bundles to the moduli of holomorphic bundles and to the moduli of
curves. They are  extensions of the Hitchin systems by the inclusion
the moduli of curves.
In contrast with the Hitchin systems the algebra
of integrals are noncommutative on GHS. We discuss the quantization
procedure in our example. The quantization of the quadratic integral leads to
 the KZB equation. We present the explicite form of higher
 quantum Hitchin integrals,
 which upon on reducing from GHS phase space to the Hitchin phase space
 gives a particular example of the Belinson-Drinfeld commutative algebra of
 differential operators on the moduli of holomorphic bundles.
\end{abstract}

\section{Introduction}
The  Knizhnik-Zamolodchikov system of equations \cite{KZ}
relates the dependence
of conformal blocks  in the Wess-Zumino-Witten-Novikov (WZWN) theory
on the positions of the insertion points on a sphere. Later this system
 was generalized on
higher genera curves \cite{B}. In the generic situation we will call it
 Knizhnik-Zamolodchikov-Bernard (KZB) system.
The further progress in the investigation of KZB
was achieved in \cite{L,I}.
Generically it
describes the dependence of conformal blocks in WZWN theory on the moduli of
 curves with
marked points.
Another type of independent variables in  the  KZB system is the moduli
of holomorphic bundles on a curve $\Si$.
The connection between
the moduli of  curves and the moduli of
bundles brings together by the Sugawara construction,
which is a starting point in
the derivation of KZB.

KZB can be derived in a completely geometric way
from the Chern-Simons theory on $\Si\times R$ \cite {H1,APW}.
The physical space
of this theory is the moduli space ${\cal M}^{fl}$
 of flat $G$ bundles over
 $\Si$,
where $G$ is a compact simple group. It is a finite-dimensional
symplectic  manifold.
 It can be endowed with a complex
structure induced from $\Si$ and therefore ${\cal M}^{fl}$
becomes a K\"{a}hler
manifold.
 There is a family
of K\"{a}hler structures
on ${\cal M}^{fl}$ depending on a complex structure on $\Si$ we  picked up.
 Let ${\cal L}=\det\bar{\p}_A$ be the determinant line bundle
  of family of operators on
$\Si$  parametrized by $G$ connections $A$,
$k$ is the "level", $h^{\vee}$ is the dual Coxeter number of $G$.
In the geometric quantization of ${\cal M}^{fl}$ for a K\"{a}hler
polarization one takes as a physical Hilbert space the space
of global holomorphic sections of
${\cal L}^{\otimes(k+h^{\vee})}$
over ${\cal M}^{fl}$.
This space formally depends on the choice of the complex structure.
 It leads to a bundle
of quantum Hilbert spaces over the Teichm\"{u}ller
space ${\cal T}$.
 According to \cite {H1,APW}  there exists a
projective identification of different fibers which
makes the bundle projectively flat. It gives rise to
the canonical (independent on the choice of complex structures) quantization.
It turns out that this bundle is the same as
the bundle of conformal blocks in
the WZWN theory.
In these terms the KZB operators are identified with a
projectively flat connections in the bundle.
The connection has a form of a heat type operator
\beq{0}
\nabla^{KZB}=(k+h^{\vee})\frac{\p}{\p_{\tau}}+B.
\eq
Here the first term is derivation along the Teichm\"{u}ller
space
and $B$ is
the second order differential operator on ${\cal M}^{fl}$ .
Solutions to
the corresponding heat
equation  for $G=U(1)$ are the theta functions on the Jacobian.
For nonabelian gauge
group they are called   the nonabelian
theta functions .

Here we adopt a different point of view on the  KZB system (\ref{0}).
Consider the classical Hitchin system \cite{H}. It is defined as a
 classical system on the cotangent bundles $T^*{\cal M}^{hol}$
 to the moduli
of holomorphic $G^{\bf C}$ bundles to a Riemann curve. Here $G^{\bf C}$
is the complex form of $G$. It is an integrable system since
it has $(\dim {\cal M}^{hol})$ integrals in
involution. As it happens in the theory of interable systems,
the Hitchin system
is obtained from \\ "a big free system" by the Hamiltonian reduction.
  The second order Hitchin integral after
quantization coincides with the operator $B$ in (\ref{0}) and therefore it
gives rise the KZB on the critical level $k=-h^{\vee}$. Our main goal
is a continuation of this correspondence
between the KZB operators and quadratic integrals
in some quantum dynamical system beyond
the critical level. The integrals should
 depend on
$(k+h^{\vee})$ in a such way that for $k=-h^{\vee}$ they coincide with the
Hitchin Hamiltonian.
In spite of the presence of the first order derivatives in the KZB operator it
 will be identified with
 a stationary Schr\"{o}dinger operator and conformal
blocks will be described as the ground states wave functions.

 We start from a classical Hamiltonian
picture based on some generalization of the Hitchin system.
We extend this approach by the introducing the dynamics along
the Teichm\"{u}ller
space ${\cal T}$ as well.
The cotangent bundle to the both type of variables\\
$T^*({\cal M}^{hol}\times{\cal T}) $
is obtained by the
Hamiltonian reduction from  "the big phase space", which is an extension of
the Hitchin phase space by including parameters of curves . The
gauge group acting on it is the semidirect product of the diffeomorphisms
of curve and
the gauge group of the bundle. There is a set of invariant Hamiltonians
on this phase space.
We will call the reduced system the Generalized Hitchin system (GHS).
 We push down the quadratic Hamiltonian from the big phase space to
the phase space of GHS and quantize it.
The Schr\"{o}dinger operator
take the parabolic form. It can be identify
with the KZB operator up to a fixing some
parameters. The role of later is still unclear and this is one of reasons
which does not allow to identify
 the
both kinds of constructions completely. Another reason is that,
 strictly speaking,
the quantization procedure is
incomplete. Since the symplectic form of our system is of the $(2,0)$ type
the direct application of the geometric quantization does not work.
 Therefore
we did not succeed to define
 the physical Hilbert space .

Our construction is closed to \cite {APW}. In this work the physical phase
space  ${\cal M}^{fl}$ is constructed as the symplectic
 quotient of the space
of $G$ connections
on $\Si$ by the group of bundle automorphisms. The Teichm\"{u}ller space
is included in the game on the next stage, as it was described above.
In our construction the physical phase space
$T^*({\cal M}^{hol}\times{\cal T}) $
is also the symplectic quotient of cotangent bundle to
the set of curves and holomorphic bundles over them.
The gauge group is the semidirect
product of diffeomorphisms of curves and holomorphic bundles automorphisms.
Since for stable bundles there exists a map from
${\cal M}^{fl}$ to ${\cal M}^{hol}$ (Narasimhan-Seshardi theorem)
our physical phase space being
projected on the $T^*{\cal M}^{hol}$
component is similar to the cotangent bundle to ${\cal M}^{fl}$.
The main difference is the consideration of the moduli of bundles and the
Teichm\"{u}ller space  on  equal terms.

We use in our construction  of the moduli of holomorphic bundles the
Schottky\\ parametrization
of curves. It leads to the \v{C}ech like description of the
 moduli space, while the original construction
\cite{H} is based on the Dolbeault cohomologies.
 At the moment this difference
is looked rather technical and leads eventually
to the same result, but it may
have some advantages in the quantization
of generic systems which is postponed
for future.
Here we restrict ourselves by the simplest nontrivial
example - an elliptic curve with a marked point. The KZB equation in
this case has been written explicitly
  in \cite{FG,EK}. Its derivation is the direct application
  of the Sugawara construction  for the representation of
level $k$ of the  loop groups .
It takes the form of the heat equation with the additional elliptic Calogero
potential. The appearing of well known integrable systems
 in the Hitchin approach is not a new phenomena \cite{M,Ne,ER}.
 Among them is $N$-body elliptic Calogero system. Though, due to the
exploiting the Schottky parameters, our derivation
 is different from \cite{Ne}, where it has been derived,
we obtain
the same system with the
 additional degree of freedom along
the module of elliptic curves as it should be.

We also consider the higher quantum integrals. Upon reducing them to the
conventional Hitchin subspace they leads
to a commuting algebra of differential
operators on the moduli of holomorphic bundles over elliptic curves with
a marked point. It is particular example of the Beilinson-Drinfeld algebra
\cite{BD}. The later is defined on the moduli of holomorphic bundles over
curves of genus $g>1$ as a result of quantization of the Hitchin system.

\section{Classical system}
Let
$$\Si_{\tau}={\bf C}^*/q^{\bf Z},~q=\exp 2\pi i\tau,~$$
be a family of elliptic curves.
The holomorphic principal $GL(N,{\bf C})$ bundle $P$ over $\Si_{\tau}$
with  sections $h(z)\in\Om^0(\Si_{\tau}, P)$
can be define by the
transition map $g(z)\in L(GL(N,{\bf C}))$
$$h(z)=g(z)h(qz).$$
Two bundles are equivalent if their transition
maps are conjugated in the following sense
\beq{1}
g_1(z)=f(z)g(z)f^{-1}(qz).
\eq
This action defines the gauge group $\{f(z)\}$, which is also
$L(GL(N,{\bf C}))$.
The last relation suggests that instead of $\{g(z)\}\sim L(GL(N,{\bf C}))$
we should consider the semidirect product
$${\cal G}=\ti {L}( GL(N,{\bf C}))=\{(\exp(2\pi i\tau z\p,g(z))\},$$
where $\p=\p_z$. The group element can be represented as the product
$g(z)T_q,~q=\exp(2\pi i\tau )$, where $T_q$ is the shift operator $T_qg(z)=
g(qz)$
 with the evident multiplication
$g_1(z)T_{q_1}\cdot g_2(z)T_{q_2}=g_1(z)g_2(q_1z)T_{q_1q_2}$.
Note that the adjoint action of $L(GL(N,{\bf C}))$ preserves $q$.
 We have Lie$({\cal G})=\{2\pi i\tau z\p+x(z),~
x(z)\in\Om^{0}(S^1,End P)\}$. The dual space
 Lie$^*({\cal G})=$
$\{\frac{\xi}{2\pi i}z^{-2}d^2
+ \phi(z),~\xi\in{\bf C},~\phi(z)\in\Om^{1}(S^1,End P)\}$
is defined by the pairing
$\xi\tau+\frac{1}{2\pi i}\tr\int_{S^1}\phi(z)x(z)$.

Consider the cotangent bundle $T^*{\cal G}$ .
The gauge transformation (\ref {1}) can be lifted on $T^*{\cal G}$
$$
x(z)\rar f(z)x(z)f^{-1}(z)+2\pi i\tau f(z)z\p f^{-1}(z),
$$
\beq{2}
\phi\rar f(z)\phi(z)f^{-1}(z),~~\tau\rar\tau,
\eq
$$\xi\rar \xi-\frac{1}{2\pi i}Res [z\tr (dff^{-1}(z)\phi(z))].$$
In addition we have a marked point on $\Si_{\tau}$ at $z=1$.
Classically we can put
in it additional degrees of freedom in the same way as it was
done in (\cite{Ne,GN,GH,Kr}). They are
defined by a symplectic manifold which is a cotangent bundle to
$G=GL(N,{\bf C})$.
In other words it is the pair
$p\in {\rm Lie}^*(GL(N,{\bf C})),h\in GL(N,{\bf C})$.
The gauge transform $f(z)$ acts  as the evaluation map at $z=1$
$p\rar f(1)p f^{-1}(1)$ and $h\rar f(1)h$.

Therefore we have the  manifold with fields
$ {\cal R}'=\{[T^*{\cal G}=(\xi,\tau,\phi(z),g(z))]
\cup \{T^*G=(p,h)\}.$
We endow it with the standard symplectic structure.
 By means of the Maurer-Cartan form on
 ${\cal G}~~Y(\tau,g)=2\pi iD\tau z\p+ g^{-1}Dg(q^{-1}z)$
we can define the canonical Liouville two form on  the cotangent bundle
$T^*{\cal G}$ . Adding the same form on $T^*G$  we obtain
the symplectic form which takes in account the contribution from
 the marked point
\beq{3}
\om=D\xi D\tau+\frac{1}{2\pi i}\tr\int_{S^1}D(\phi(qz),g^{-1}Dg(z)) +
D\tr (p h^{-1}Dh).
\eq
It is invariant under the gauge action of $L(GL(N,{\bf C}))$.
There is  the set of $N+1$ independent invariant
Hamiltonians. They have the following form. Represent the element $g(z)$ as
$g(z)=\exp(2\pi i\tau z\p+x(z))T_q^{-1}$. Then the following quantities are
invariant under the gauge action (\ref{2})
\beq{4}
I_n=\frac{\tau^{n-1}}{2\pi in}\tr\int_{S^1}\phi^n(z),~n=1,\ldots,N,
\eq
$$L=\xi\tau+\frac{1}{2\pi i}\tr\int_{S^1}(\phi(z)x(z)).$$
Hamiltonians $I_n$ are the same as in the Hitchin system \cite{H}. They
define   motions along the
moduli of the bundle. The additional Hamiltonian $L$  includes
also a motion along the
elliptic module in the consistent way. While $I_n$ are well defined on the
moduli of holomorphic bundles, $L$ is defined only on the covering
 of the both kind
of moduli.
The Hamiltonians form the closed noncommutative algebra with respect
the Poisson brackets
coming from (\ref{3})
\beq{4a}
\{I_j,I_l\}=0,~\{L,I_n\}=-I_n.
\eq
The algebra is defined also only on the covering of the moduli space.
The gauge action on $ {\cal R}'$ of $L(GL(N,{\bf C}))$ produces
the moment map
$\mu :{\cal R}'\rar{\rm
Lie }^*(L(GL(N,{\bf C})))$,
which we put equal to zero
\beq{5}
\mu=g(z)\phi(qz)g^{-1}(z)-\phi(z)+hph^{-1}\de(z)=0,
\eq
where $\de(z)$ is defined as $\frac{1}{2\pi i}\int_{S^1}F(z)\de(z)=F(1)$.
For  higher genera curves $\Si$ it is possible to add to the gauge
transform the action of the diffeomorphisms or by what they are replaced in
specific realizations of curves. They produce the additional
moment constraints which lie in ${\rm Lie^*(Diff}\Si)$. In the present
construction this action
 has been fixed already, since we have the forth mentioned realization of
the elliptic curves as an annulus with the parameter $\tau$.

It is instructive to write down the  action corresponding
to the theory defined by
the symplectic form (\ref{3}), constraints (\ref{5}) and
a Hamiltonian that can
be an arbitrary linear combination of the invariant Hamiltonians (\ref{4})
\beq{6a}
S=\int_tdt[\xi\p_t\tau+\tr(ph^{-1}\p_th)+
\eq
$$
\frac{1}{2\pi i}\tr\int_{S^1}(\phi(qz)A_0(z)^{g(z)}-
\phi(z)A_0(z)-\de(z)hph^{-1}(z)A_0(z))
-H(L,I_1,\ldots I_N)].
$$
Here $A_0(z)$ is the Lagrange multiplier, $A_0^{g}=g^{-1}\p_tg+
g^{-1}A_0g$. The action describes nonlocal theory
on the cylinder $S^1\times R$.
The nonlocality is a result of the Schottky parametrization.
The Hitchin approach based on the Dolbeault cohomologies leads to a local
theory but in $2+1$ dimension \cite{GN1}. Our
description seems to be more preferable in quantum calculations.

We will construct the reduced space ${\cal R}={\cal R}'//L(GL(N,{\bf C}))=
\mu^{-1}(0)/L(GL(N,{\bf C}))$. By the gauge transform (\ref{2})
 we can diagonalize
 $g(z)$.
 Moreover it can be chosen as a constant matrix since the vector bundles
over elliptic curves are reduced to a sum of linear bundles.
Let represent it as
 $f(z)g(z)f^{-1}(qz)=\di\exp(2\pi i\vec{u}),~\vec{u}=(u_1,\ldots,u_N)
 \in {\cal H}$,
where ${\cal H}$ is the Cartan subalgebra of Lie$(GL(N,{\bf C}))$.
The remnant gauge symmetries that preserves the diagonal
 $z$-independent form
of matrix $\di\exp(2\pi i\vec{u})$ will be discussed at the
end of this section.
The gauge invariant subvariety in $T^*G$ at the marked point is the coadjoint
orbit, which is fixed by the invariants
 ${\cal O}_p=\{\chi=hph^{-1}|\tr p^j=c_j\}$.
 We still have a freedom to act by the diagonal gauge transformations
$({\bf C}^*)^N$ on the points of the orbit ${\cal O}_p$. This action
is not free   - only $({\bf C}^*)^{N-l}$ part acts effectively,
where $l+1$ is the number non equal eigenvalues of $p$.
 Thus the reduced space
is
$${\cal R}=T^*{\cal N},({\cal O}_p//({\bf C}^*)^N),$$
where $T^*{\cal N}=T^*{\cal G}//L(GL(N,{\bf C}))$ is the
cotangent bundle to the moduli of elliptic curves and the moduli
of holomorphic
bundles. The space ${\cal O}_p//({\bf C}^*)^N$ defines the parabolic
structure at the marked point $z=1$. The dimension of ${\cal R}$ is equal
$\dim{\cal R}=2N+2+[\dim{\cal O}_p-2(N-l)].$ Here the first term is
responsible
for the holomorphic bundles moduli, the second for
the elliptic module and the
last comes from the orbit in the marked point.

After the resolving the moment constraints (\ref{5})
$$\phi_{j,j}(qz)-\phi_{j,j}(z)=\chi_{j,j}\de(z),~(\chi=hph^{-1}),$$
$$\phi_{j,k}(qz)\exp 2\pi i(u_j-u_k)-\phi_{j,k}(z)=\chi_{j,k}\de(z)$$
 we find
$$\phi_{j,j}(z)=p_j,~\chi_{j,j}=0,$$
$$\phi_{j,k}=
-\frac
{\chi_{j,k}}{2\pi i }
\frac
{\te(u_j-u_k-\ze)\te'(0)}
{\te(u_j-u_k)\te(\ze)},~z=\exp 2\pi i \ze$$
where $p_j,j=1,\ldots ,N$ are new free parameters and
$\te(\ze)=\sum_{n\in {\bf Z}}e^{\pi i(n^2\tau+2n\ze)}$.
The symplectic form (\ref{3}) on the reduced space takes the form
\beq{6}
\om^{red}=D\xi D\tau+D\vec{p}\cdot D\vec{u} +
\tr D( h^{-1}\chi Dh).
\eq

Consider the quadratic Hamiltonian which is the linear combination
$H=\al L+I_2, ~\al\in {\bf C}$. Its form resembles the Sugawara
construction.  We can also consider higher Hamiltonians as well.
 After the reduction $H$ takes
the form of the N-body elliptic Calogero Hamiltonian
with the spins \cite{GN,GH,Kr} and the additional terms coming from $L$
\beq{7}
H=\al(\xi\tau+\vec{p}\cdot \vec{u})+\frac{\tau}{2}(\vec{p}\cdot\vec{p}+
\sum_{j>k}^N
\chi_{j,k}\chi_{k,j}U(u_j-u_k,\tau)
\eq
$$
U(u,\tau)=\frac{1}{4\pi^2}
[\wp(u|\tau)+\frac{\pi^2}{3}E_2(\tau)]).
$$
Here  $\wp(u|\tau)$ is the Weierstrass elliptic function and $E_2(\tau)$
is the normalized Eisenstein series which is defined as \cite{Kob}
$$E_2(\tau)=
\frac{3}{\pi^2}\sum_{m=-\infty}^{\infty}\sum_{n=-\infty}^{\infty '}
\frac{1}{(m\tau+n)^2}=\frac{24}{2\pi i}\frac{\eta'(\tau)}{\eta(\tau)},$$
where ' means that $n\neq 0$ if $m=0$ and $\eta(\tau)$ is the Dedekind
$\eta$-function . The spin degrees of freedom $\chi$ can be described
in a more explicit form (see details
in \cite{Ne}). Consider the flag variety corresponding to the orbit
${\cal O}_p$
$V^0=C^N\supset V^1\supset\ldots V^l$. Then $\chi_{j,k}\chi_{k,j}$ can be
replaced by $\tr_{V^1}S_jS_k$, where $S_j,S_k$ are matrices in $V^1$.

The symmetries of the Hamiltonian are simultaneous permutations of $u_k$ and
$p_k$ extended by the shifts
 \beq{8}
\vec{u}\rar \vec{u}+n\vec{\be}\tau ,~p_i\rar p_j,~\tau\rar\tau,
{}~\xi\rar\xi -n (\vec{\be},\vec{p}),
\eq
$$n\in{\bf Z},~~\vec{\be}\in R^{\vee}({\rm dual~root~system}),$$
 The symmetries of the symplectic form can be extended
 by the additional real shifts
of $\vec{u}$: $\vec{u}\rar \vec{u}+m\vec{\ga},~\vec{\ga}\in R^{\vee}$.
The group $\hat{W}$ which is the semidirect
extension of the  Weyl group $W$ (the permutations at hand) by shifts
${\bf Z}R^{\vee}\tau+{\bf Z}R^{\vee}$ is a particular example of
complex crystallographic Coxeter groups \cite{BS}. The factor
${\cal H}/\hat{W}$ decribes almost all holomorphic $GL(N,\bf C)$ bundles
 over $\Si_{\tau}$ (see, for example \cite{EFK}).
  Thus the symplectic structure is well
define on ${\cal M}^{hol}\times{\cal T}$. Let $W'$ be the extension
of $W$ by the part of
the shifts ${\bf Z}R^{\vee}\tau$. In other words
$\hat{W}=W'\oslash {\bf Z}R^{\vee}$. Consider the
covering   ${\cal M}^{'hol}={\cal H}/W'$ of  ${\cal M}^{hol}$.
The Hamiltonian (\ref{7}) due to the second
term lives on the covering ${\cal M}^{'hol}\times{\cal T}$.
We will see in the next section
 that after the quantization that Hamiltonian can be push down on
 ${\cal M}^{hol}\times{\cal T}$ as well.

\section {Quantum system}

We should quantize the classical Hamiltonian system with  symplectic
structure (\ref{6}) and  Hamiltonian (\ref{7}) and with additional
symmetries (\ref{8}). There is also the algebra
of higher Hamiltonians (\ref{4a})
which is desirable to preserve on the quantum level.

There are two type of variables - those that related to the coorbit
 and  to the moduli spaces.
The quantization of coadjoint orbits is well developed procedure  \cite{Ko}.
 The symplectic form on the orbit ${\cal O}_p$ is the Kirillov-Kostant form.
Moreover  coadjoint orbits are complex manifolds such that the symplectic
form has type $(1,1)$ and so it allows to use
 the K\"{a}hler polarization after the holomorphic prequantization. The
only restriction is the integrability of the orbit ${\cal O}_p$
which gives rise
to the correspondence of $p$ to a dominant weight $\La_p$.
 After the quantization
we obtain the representation of the dominant weight $\La_p$. Functions
 on the orbit become operators in the representation space.
The factorization of ${\cal O}_J$ under the action
of the diagonal subgroup leads
on the quantum level to the additional restriction -
only vectors from the zero weight subspace contributes
to the quantum Hilbert state.

The quantization of others degrees of freedom is a rather subtle procedure.
Note first of all that the symplectic structure (\ref{6}) is the holomorphic
$(2,0)$ form on $T^*{\cal N}$ . This property follows from our definition of
the phase space as the cotangent bundle to the complex manifold.
This form defines the polarization from the very beginning.
But in contrast with the $(1,1)$ forms
it does not lead to topological restrictions on the physical Hilbert space
since it defines only trivial line bundles. Another  unpleasant property
of the system is the definition of the "coordinate type" variables $\tau$ and
$\vec u$ which do not belong to the moduli spaces, but only their
coverings. As we have mentioned above it comes from our definition of the
hamiltonian $L$. We will see later that the quantization improves partly
the situation - the quantum Hamiltonian is well defined on the moduli of
holomorphic bundles.

 There are evidences that the first difficulty can be overcome as well
at least
 in the original Hitchin systems in which we encounter with the same
problem of the quantization of $(2,0)$ form.

 A.Beilinson and V.Drinfeld \cite{BD} proposed
 the quantization of the original Hitchin system,
defined on the cotangent bundles to the moduli space of holomorphic
$G$-bundles  over arbitrary curves $\Si$ of genus
$g>1$ without marked points. They constructed  a canonical morphisms from
the ring of polynomial functions generated by the Hitchin integrals of  type
(\ref{4a}) to a sheaf of differential operators on the moduli of holomorphic
bundles ${\cal M}_G(\Si)$. The algebra of differential
operators is commutative and globally defined on the moduli space.
Let $K$ be a canonical bundle on ${\cal M}_G(\Si)$.
It is defined as the tensor degree of the determinant line bundle
${\cal L}=\det\bar{\p}_A,~~$
$K={\cal L}^{\otimes h^{\vee}}$, where $h^{\vee}$ is the dual Coxeter number.
Then the physical Hilbert space is the space of section of the bundle
  $K^{1/2}$ of half-forms. They are generated by the center
${\cal C}_{-h^{\vee}}$
of the universal enveloping algebra $U_k({\rm Lie }\hat{LG})$
on the critical level
$k=-h^{\vee}$. It was proved in $\cite{FF}$ that  ${\cal C}_{-h^{\vee}}$ is
isomorphic to the classical $W_{G'}$ for the dual group $G'$.
This construction
eventually leads to a consistent system of differential equations
on ${\cal M}_G(\Si)$. The second order differential operators
coincide with the
 KZB operators  depending on the moduli of holomorphic bundles.
 Thus one way to derive the full KZB system as the
Hamiltonian system is the
reproducing the Beilinson-Drinfeld procedure for the space unifying
the both kind of moduli. Note that essential part of their construction
is a representation of the
moduli of holomorphic bundles as a double coset space
of the loop group $L(G)$.
 On the other hand
the moduli of curves is also can be represented as a double coset space
of the diffeomorphisms of a circle.
Their unification does not look very natural. It seems that
 a more relevant construction
is the unification of
the moduli of holomorphic $G$-bundles with the moduli of corresponding $W_G$
 structures.
In contrast with the original
Hitchin system the family of integrals are noncommutative on the
unified  space, as we have seen in
our simplest example (\ref{4a}). Of course noncommutativity
 holds in a generic case.

We stop here the discussion about the rigorous quantization of the GH system.
 We shall not attempt to define a  Hilbert space in which our observables act
 and restrict ourselves to the consideration of a formal algebra of operators.
Since the quantization of degrees related to the orbit is straightforward
 we will consider the system with a simplest nontrivial orbit.
The classical Hamiltonian  in this case is
$$
H_{\al}=\al(\xi\tau+\vec{p}\cdot \vec{u})+\frac{\tau}{2}(\vec{p}\cdot\vec{p}
+\nu^2
U(\vec{u}|\tau)),
$$
where we have chosen the most degenerated orbit ${\bf CP}^{N-1}$.
 It is conjugate to the matrix
 $\nu (e^T\otimes e-\di(1,\dots,1)),~e=(1,\ldots,1)$ .
  Define the following quantization of the classical quantities
$$\rho(u_j)=u_j,~\rho(p_j)=\nabla_j=i\p_j+\frac{i\al u_j}{\tau},
{}~(\p_j=\frac{\p}{\p_{u_j}})
$$
\beq{9c}
\rho(\tau)=\tau,~
\rho({\xi})=\nabla_{\tau}=i\p_\tau-\frac{i\al (\vec{u}\cdot\vec{u})}{2\tau^2}.
\eq
 These covariant derivatives
can be gauge transformed from the canonical operators  by the Gaussian twist
$\Pi=\exp\frac
{i\al (\vec{u},\vec{u})}
{2\tau}$ . Namely,
$$\nabla_j=\Pi^{-1}\circ i\p_j\circ\Pi,~~
\nabla_{\tau}=\Pi^{-1}\circ i\p_\tau\circ\Pi.$$
Thus (\ref{9c}) is the quantum counterpart of the classical algebra of
observables (\ref{6}).
The quantization leads to the
quantum Hamiltonian
$$
\hat{H}_{\al}=\al(\tau\nabla_{\tau}+\sum_{j=1}^N u_j\nabla_j)+
\frac{\tau}{2}(\sum_{j=1}^N \nabla_j^2+\nu^2
U(\vec{u}|\tau)),
$$
where $\nu^2=\frac{m(m+1)}{2},~m\in{\bf N}$ due to the quantization of the
orbit.
We study the eigenvalue problem
$$
\hat{H}_{\al}\Psi_E(\vec{u},\tau)=E\Psi_E(\vec{u},\tau),
$$
where
$\Psi_E(\vec{u},\tau)$ is defined on the product of
$\{\vec{u}|{\cal M}^{'hol}={\cal H}/W'\sim
{\bf C}^N/{\bf Z}R^{\vee}\tau\}$
 and the halfplane $Im\tau>0$.
After some algebra it can be transformed to
$$
[i\al\p_{\tau}+
\frac{1}{2}(\sum_{j=1}^N -\p^2_j+\nu^2U(\vec{u}|\tau))]\Psi_E(\vec{u},\tau)
=E'\Psi_E(\vec{u},\tau),
$$
where
$E'=\frac{1}{\tau}(E-i\al N/2)$.
Now the quantum Hamiltonian $\hat{H}_{\al}$ in contrast with
classical one is well defined on the moduli of bundle, since it is
invariant under the shift $\vec{u}\rar \vec{u}+{\bf Z}R^{\vee} $.
Therefore it lives
on  ${\cal H}/\hat{W}$, which
defines the moduli of the $GL(N,{\bf C})$ bundles over the elliptic curve.
The Hamiltonian $\hat{H}_{\al}$ for $i\al=k+h^{\vee} (h^{\vee}=N)$
after the gauge transform
${\cal D}^{-1}\hat{H}_{\al}{\cal D},~({\cal D}$ is the Weyl-Kac denominator)
 coincides
 with the KZB operator (\ref{0}) for the elliptic
curve with a marked point \cite{FG,EK}.
 Gauge transformed groundstate wave functions
 ${\cal D}^{-1}\Psi_{i\al N/2}(\vec{u},\tau)$
  gives the conformal blocks.
They are sections of the line
bundle ${\cal L}^{k+h^{\vee}}$ over ${\cal H}/\hat{W}$.
Recently they  were analized
in \cite{FW,FG1}.
 For $\nu=0, E'=0$ we come to the standard heat
equations, which solutions are the level $i\al$ theta functions on
${\bf C}^N/{\bf Z}R^{\vee}+{\bf Z}R^{\vee}\tau.$

 Though the higher hamiltonians (\ref{4}) are beyond the problems around
 the KZB system
they can be quantized as well. It can be done by the using old results
 \cite{OP} about
quantum Calogero systems. The corrections
to the integrals independing on $\vec{u}$ and depending on $\tau$ can be done
in such way that the whole algebra (\ref{4a}) will be satisfied
for commutators
on ${\cal M}^{'hol}\times{\cal T}$.

For the quantization it is more convinient to consider another basis in
the integrals instead of $\hat{I}_n$. Let $\hat{J}_n$ be operators
which  symbols
    are symmetric polynomials of the form $p_1p_2\ldots p_n+$ permutations.
The reason is that $\hat{J}_n$ as well as their commutators contain only
well defined terms with commuting multipliers.
It is easy to show that  the following reccurence  relation holds
$$i\hat{J}_{n-1}=\frac{1}{N-n+1}[\sum_{j=1}^Nu_j,\hat{J_n}].$$
On the other hand the highest integral has the following form
$$\hat{J}_N=\exp\{-\frac{\nu}{2}\sum_{k,l}x^2(u_k-u_l)\p_{\hat{p}_k}
\p_{\hat{p}_l}\}
\hat{p}_1\ldots\hat{p}_N,$$
where $\hat{p}_k=\p_{u_k}$ and $x(u)=\frac{\wp(1/2)-\wp(\tau/2)}
{\sqrt{\wp(u)-\wp(\tau/2)}}+c(\tau).$

There is also possible to write down the explicit form of the lowest integral
 $\hat{I}_n$. Intruduce
the following notations $x_{k,l}=x(u_k-u_l),~<x_{k,l}>$ is the trace of the
matrix $(x_{k,l})$. Then
$$\hat{I}_3=\sum_{k=1}^N\hat{p}^3_k+3\nu^2\sum_{k\neq l}x^2_{k,l}\hat{p}_l,$$
$$\hat{I}_4=\sum_{k=1}^N\hat{p}^4_k+2\nu^2\sum_{k\neq l}x^2_{k,l}(2\hat{p}_l^2+
\hat{p}_k\hat{p}_l)+\nu^4<x^4_{k,l}>$$
$$+\nu^2\sum_{k\neq l}[2(x^2_{k,l})'i\hat{p_l}-(x^2_{k,l})''],$$
$$\hat{I}_5=\sum_{k=1}^N\hat{p}^5_k+5\nu^2\sum_{k\neq l}x^2_{k,l}(2\hat{p}_l^3+
\hat{p}^2_k\hat{p}_l)+5\nu^5<x^5_{k,l}\di (\hat{p}_1,\ldots ,\hat{p}_N)>$$
$$5\nu^2\sum_{k\neq l}[2(x^2_{k,l})'i\hat{p_l}-(x^2_{k,l})''\hat{p}_l].$$
The last lines in the expressions for $\hat{I}_4$ and $\hat{I}_5$ are quantum
corrections to the classical integrals.

These formulae give the explicit form of the Belinson-Drinfeld commutative
algebra of global differential operators on the moduli of holomorphic bundles
over the elliptic curve with a marked point.

To conclude, we established the connection between the KZB equation on
an elliptic curve  with a marked point and the heat type extension of the
elliptic Calogero quantum Hamiltonian. The later was constructed by means of
the Hamiltonian reduction and lives on the moduli of holomorphic bundles
and the  Teichm\"{u}ller
space. The description of the system is suffered
from the absence of correct definition of the quantum Hilbert space. It
seems that the construction can be improved by including in the game
from the very beginning on the classical level
the central charge of the loop group . In this case the classical Hamiltonians
  can be defined on the cotangent
bundle to the holomorphic moduli directly before the quantization. Apparently,
it can help to define the quantum Hilbert space as well.

\bf Acknowledgments.
{\sl I would like to thank my colleagues V.Fock, A.Gerasimov, A.Gorsky,
D.Ivanov,
A.Levin
A.Morozov, N.Nekrasov, A.Rosly  and V.Rubtsov for illuminatig discussions.
This work was prepared partly during my visits to the
the Cornell University and SISSA (Trieste).
 I am grateful to  Andre Leclair  and Loriano Bonora for the hospitality.}

\small{

}

\begin{thebibliography}{40}
\bibitem{KZ}\ V.Knizhnik and A.Zamolodchikov, Nucl.Phys. {\bf B247} (1984) 83
\bibitem{B}\ D.Bernard, Nucl.Phys. {\bf B303} (1988) 77;
 Nucl.Phys. {\bf 309} (1988) 145
\bibitem{L}\ A.Losev, {\em Coset construction and Bernard equation,} Preprint
CERN-TH.6215/91
\bibitem{I}\ D.Ivanov, {\em KZB equations on Riemann surfaces}, hep-th/9410091
 \bibitem{H1}\ Hitchin N., Flat connections and geometric quantization,
 Comm.Math.Phys., {\bf 131} (1990) 347-380
\bibitem{APW}\ Axelrod S., Della Pietra S. and Witten E.,
{\em Geometric quantization of the Chern-Simons gauge theory},
Journ. Diff. Geom., {\bf 33} (1991) 787-902
\bibitem{H}\ Hitchin N., {\em Stable  bundles and Integrable Systems},
Duke Math. Journ., {\bf 54} (1987) 91-114
\bibitem{FG}\ Falceto, F. and Gawedzky, K.,
{\em Chern-Simons states in genus 1},
Comm.Math.Phys., {\bf 159} (1994) 471-503
 \bibitem{EK}\ Etingof P. and Kirillov A., {\em Representations of affine Lie
algebras, paraboloc differential equations and L\'{a}me functions},
Duke Math. J. {\bf 74} (1994), no. 3, 585-614
\bibitem{M}\ Markman E., {\em Spectral curves and integrable systems},
 Comp. Math.,
{\bf 93} (1994) 255-290
\bibitem{Ne}\ Nekrasov N., {\em Holomorphic bundles and many-body systems},
 PUPT-1534,
 hep-th/9503157
\bibitem{ER}\ Enriques B. and Rubtsov V.,
{\em Hitchin systems, higher Gaudin operators
and r-matrices}, preprint (1995)
\bibitem{BD}\ Beilinson A. and Drinfeld V.,
 {\em Quantization of Hitchin's fibration and
and Langlands program}, preprint (1994)
\bibitem{GN1}\ Gorsky, A. and Nekrasov N., {\em Elliptic
Calogero-Moser system
 from the two dimensional current algebra},
preprint ITEP-NG/1-94, hep-th/9401021
\bibitem{GN}\ Gorsky, A. and Nekrasov N., Nucl.Phys., {\bf B414} (1994) 213;
{\bf B436} (1995) 582
\bibitem{GH}\ Gibbons J. and Hermsen T., {\em A generalization of
Calogero-Mozer
system}, Physica {\bf 11D} (1984) 337
\bibitem{Kr}\ Krichever I., Babelon O., Biley E. and Talon M.,
{\em Spin generalization
of the Calogero-Mozer system and the matrix KP equation}, Preprint LPTHE 94/42
\bibitem{Kob}\ Koblitz N., {\em Introduction to
Elliptic Curves and Modular Forms}
 Gradute Texts in Math., {\bf 97} Springer-Verlag (1984)
\bibitem{BS}\ Bernstein J.N. and Shvartsman O.V., Chevalley theorem for
complex crystallographic groups, Funct. Anal. Appl. {\bf 12} (1978), 308-310
 \bibitem{EFK}\ Etingof P., Frenkel I. and Kirillov A.,
 {\em Spherical functions on affine
Lie groups}, Yale preprint, hep-th/9407047
 \bibitem{Ko}\ Kostant B., Lect. Notes in Math., {\bf 170},
 Springer, Berlin, 1970
\bibitem{FF}\ Feigin B. and Frenkel E., {\em Affine Kac-Moody algebras
 at the critical level
 and Gelfand-Dikii algebras}, Intern. Journ. Mod. Phys. A, {\bf 7}, Suppl.
  {\bf 1A}
  (1992) 197-215
\bibitem{FW}\ Felder G. and Wieczerkowski C. {\em Conformal blocks on elliptic
curves and the Knizhnik-Zamolodchikov-Bernard equations},
 preprint, hep-th/9411004
\bibitem{FG1}\  Falceto, F. and Gawedzky, K.,
{\em Elliptic Wess-Zumino-Witten Model
 from Elliptic Chern-Simons Theory}, preprint, hep-th/9502161
\bibitem{OP}\ Olshanetsky, M. and Perelomov A.,{\em Quantum completely
integrable
 systems connected with semisimple Lie algebras},
 Lett. Math. Phys., {\bf 2} (1977), 7-13

\end{thebibliography}
\end{document}